# NaI(Tl+Li) scintillator as multirange energies neutron detector

D. Ponomarev [a,1], D. Filosofov [a], J. Khushvaktov [a], A. Lubashevskiy [a], I. Rozova [a], S. Rozov [a], K. Shakhov [a], Yu. Shitov [a], V. Timkin [a], E. Yakushev [a], I. Zhitnikov [a].

[a] *Laboratory of Nuclear Problems, JINR, Dubna, 141980, Russia*

 E-mail: ponom@jinr.ru.

ABSTRACT: Novel NaIL detector (5×6 inch) was investigated for its neutron detection in wide energy range. It has been found that the detector together with its known ability to detect the γ– radiation it also allows to distinguish neutron signals in three quasi-independent ways. It is sensitive to neutron fluxes on a level down to $10^{-3}$ cm$^{-2}$ s$^{-1}$. In this work intrinsic α– background and neutron detection sensitivity for the NaIL detector were obtained. Experimental data was compared with results of Geant4 Monte Carlo (MC).

---

[1] Corresponding author.

**Contents**

**1. Introduction**

Neutron detectors for a wide energy range from thermal to the MeV are one of the most look for tools in the physics. Unfortunately such wide range neutron detection remains difficult, especially for cases when low neutron flux has to be measured. Detailed knowledge of ambient neutrons is a crucial point in many experiments aiming a very low radioactive background. Fast neutrons are hard to study as usually measurement results are model dependent, either due to quenching in detectors [1,2], or due to indirect way of detection [3,4].

In this work we propose to use recently available commercial NaIL (NaI(Tl+Li)) detectors [5] as tools for simultaneous detection of ionizing radiation together with neutrons in a wide energy range due to presence of Li-6 in the scintillator. These scintillator detectors with the energy resolution at the same level as standard NaI(Tl) detectors have an excellent ability for pulse shape discrimination (PSD) between light and heavy particles. They are available in relatively large sizes; see for example description of our detector below, which ensures high detection efficiency.

**2. Experimental methods**

In our studies Saint-Gobain NaIL detector [5] with a NaI(Tl+Li) 7 kg crystal with size of 5 inches in diameter and 6 inch height was used. Natural lithium that has 7.5% of Li-6 is used for doping during the production of the crystal (per the company, its concentration is at least 1 percent per mol). The detector is equipped with a Hamamatsu R877-02 PMT. A CAEN DT5725 digitizer with the PSD firmware was used for the data acquisition [6]. In this work, the PSD tail to total method was applied with length of the total signal 750 ns and where the tail is the last 400 ns. Neutrons were detected in the following ways:

- Thermal neutron capture on Li-6: $^6Li+n_{th}\rightarrow {^4He}+{^3H}+4.79$ MeV, $\sigma$=940 b [7];
- Thermal neutron capture on iodine with the method proposed in [8] with detection of delayed $\gamma\gamma$– coincidences: $^{127}I+n_{th}\rightarrow {^{128}I^*}+6.8$ MeV, $\sigma_{th}$=6.2 b, further referred as $^{127}I$ method;
- It has been noticed that above reaction also has reasonably high cross section for epithermal neutrons (Figure 1). From 50 eV to 10 keV its integral is 153.9 b [9]. Thus in the bulk which could be unreachable by thermal neutrons, due to Li-6, the reaction is suitable for epithermal and fast neutron studies (see MC below).

NaIL detectors have yet another possible method for neutron detection that has not been used in the present work. Inelastic scattering of fast neutrons on I-127 with excitation of 57.6 keV level can be detected [10,11]. The method is applicable only for low background environment and will be used in our future studies.

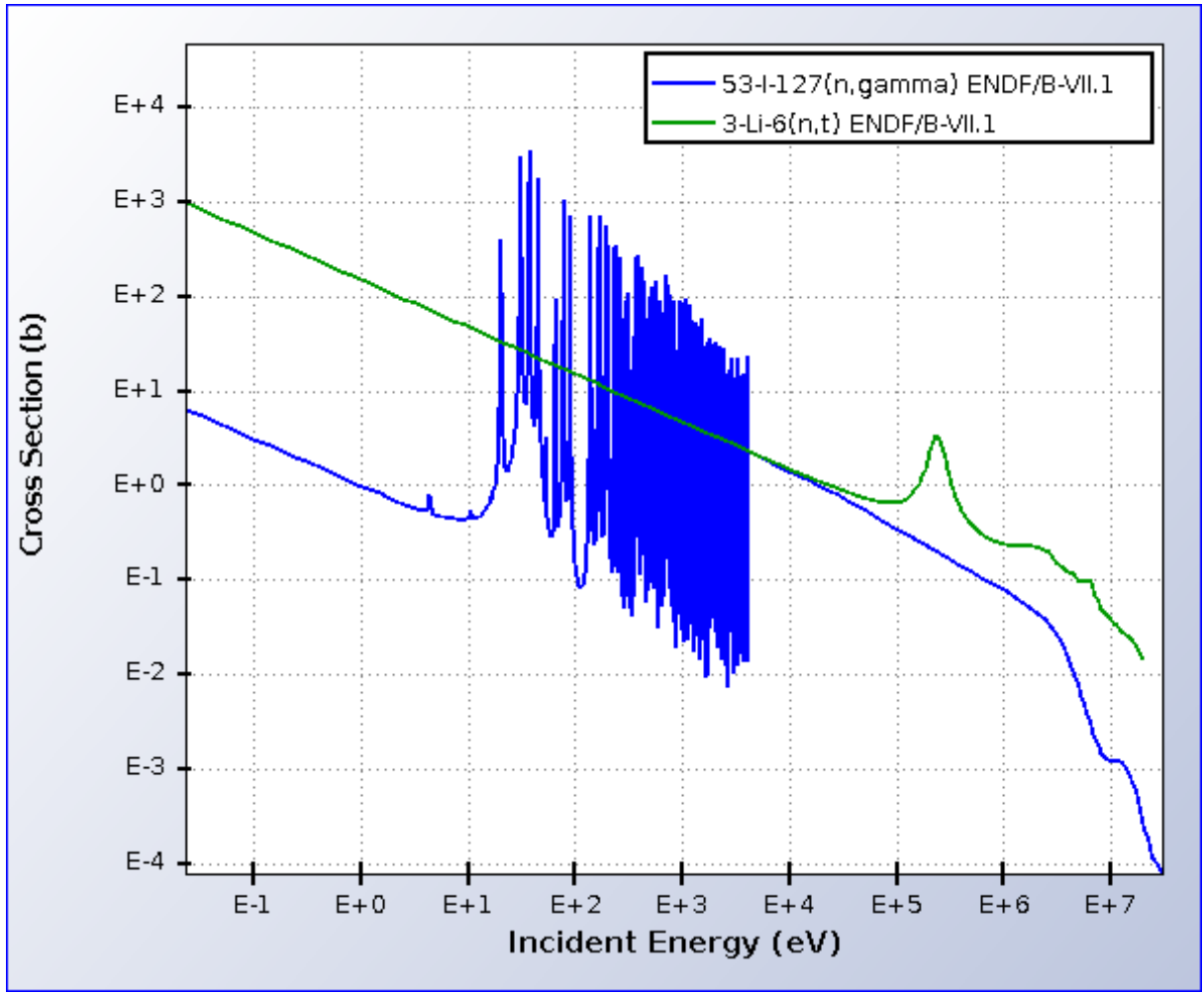

**Figure 1.** Neutron's cross sections for $^6$Li and $^{127}$I from [9].

**3. Measurements**

Preliminary check of the detector response to γ– rays was done with a Cs-137 radioactive source and with gammas of the ambient background (bare detector). For the Cs-137 γ– line 661.7 keV the energy resolution (FWHM) is 7%. Then extensive experimental tests of the NaIL detector were performed with a special shield: 10 cm of lead for moderate reduction of γ– background, covered inside with 3.3 mm layer of borated rubber (50% of boron) for suppression of thermalized neutrons. A set of measurements was performed with PuBe source ($10^4$ neutrons sec$^{-1}$) placed outside of the shield (further referred as PuBe run). Those measurements were accompanied by a run without the source with the detector in the shield (background run). Data that demonstrated the ability for neutron detection is presented in PSD diagrams: Figures 2 and 3 for the PuBe and the background runs, respectively. There is a clear island below the γ– band in Figure 2. It was identified as neutron captures on Li-6. Without the source the island remained clear (due to ambient neutrons) but its intensity becomes comparable to the intrinsic α–background (Figure 3) which is well pronounced. For a direct comparison we superimposed both diagrams in Figure 4. Some of the α– events were identified as decay of Po-214 (thanks to BiPo delayed coincidence technique) with its intensity of 170±1 counts per day (cpd). Based on that identification we were able to roughly estimate the QF (quenching factor – ratio of light yields produced by alphas and electrons) for 7.69 MeV α- particles as ~0.6.

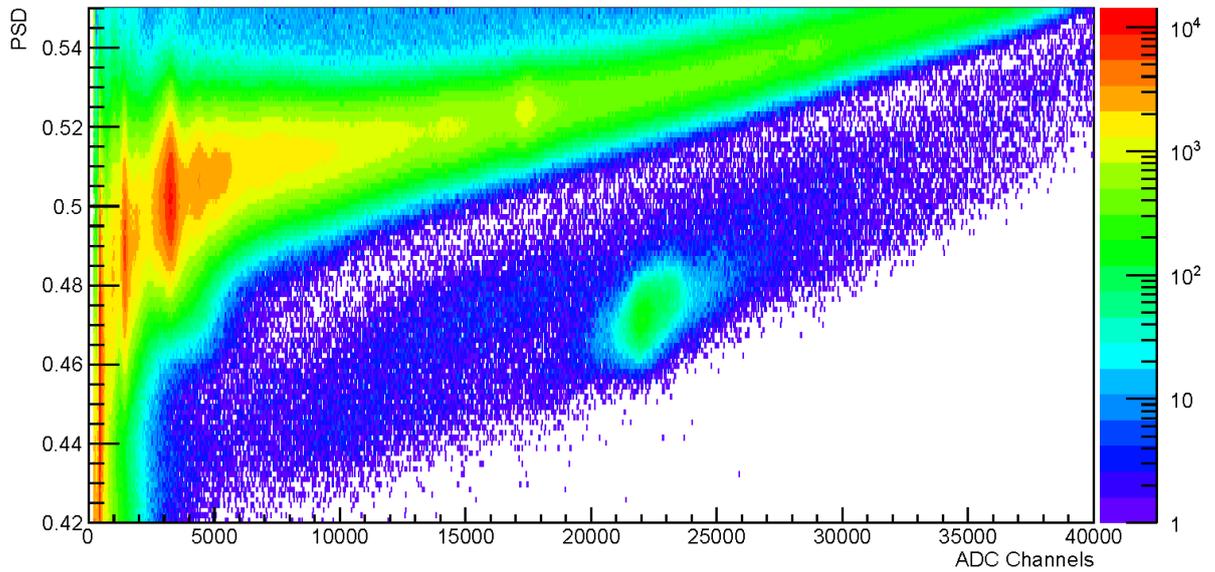

**Figure 2.** PSD diagram for the PuBe run.

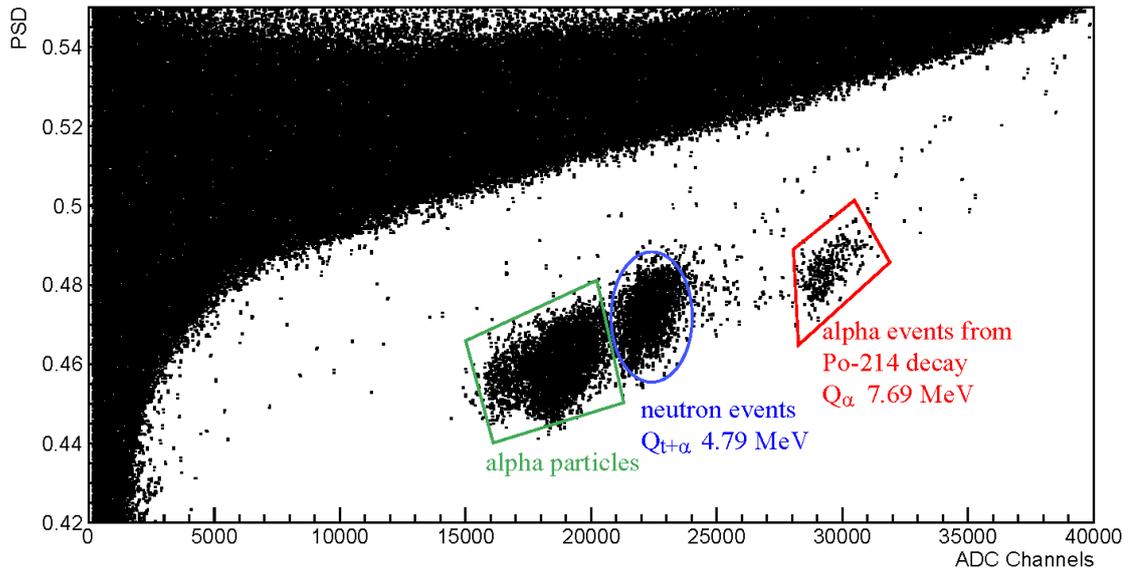

**Figure 3.** PSD diagram for the background run.

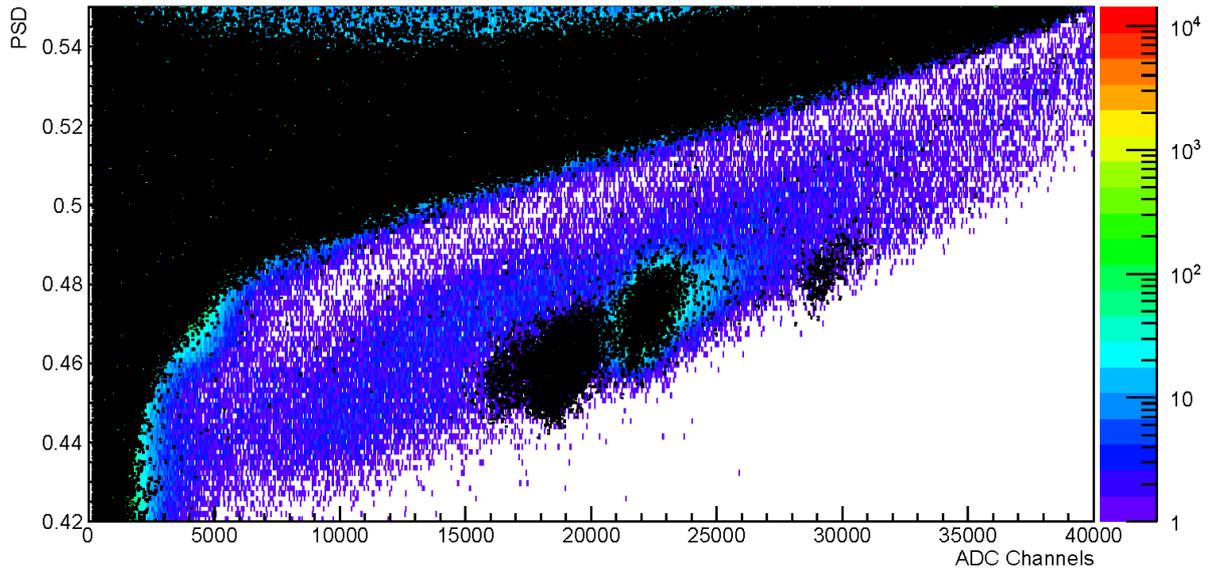

**Figure 4.** The PuBe run (color) and the background run (black) superimposed for direct comparison.

For the $^{127}$I neutron detection method the delayed time window was set to 0.9-10 μsec. The accidental background was estimated using a shifted 10.9-20 μsec delayed window. Its value for the PuBe run was 3714±61 cpd. It is clear from Figure 5, that 137.8 keV I-128 isomeric state produced by neutrons in the NaIL detector is evident. The accidental background can be compared to the observed neutron signal during the PuBe run (table 1).

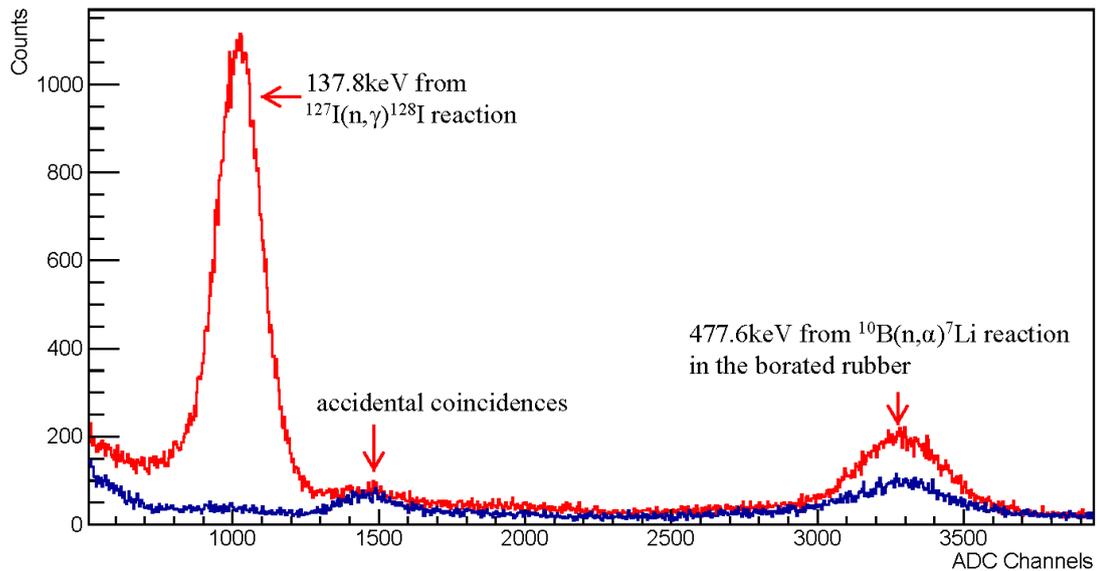

**Figure 5.** Delayed energy spectrum obtained with PuBe neutron source and the NaIL detector. Red and blue lines correspond to 0.9-10 μsec and 10.9-20 μsec delayed time windows, respectively. The peak at ~1000 ADC channels is 137.8 keV line of $^{127}$I(n,γ)$^{128}$I.

Additionally background measurements were supplemented with a run when the borated rubber was removed. Ambient neutrons were detected with both methods ($^6$Li/PSD and $^{127}$I), with results presented in Table 1.

**Table 1.** Neutron rates obtained in different runs as described in the text.

|  | Neutron rate, cpd $^{127}$I method | Neutron rate, cpd $^6$Li/PSD method | Ratio of values for columns 3 to 2 |
|---|---|---|---|
| PuBe run | 48359±283 | 95826±384 | 1.98±0.01 |
| Background, Pb+borated rubber shield | 216±6 | 523±9 | 2.4±0.1 |
| Background, Pb shield, no borated rubber | 1028±7 | 8259±6 | 8.0±0.1 |

Removing of the rubber is expected to increase thermal neutron flux inside of the shield. As result, neutron counting rate on Li-6 is expected to be higher in respect to the $^{127}$I method. Our experimental data (Table 1) is in agreement with such expectation. In presence of the PuBe source and the borated rubber shield the ratio between thermal and other neutrons is minimized. This case corresponds to the first line of Table 1, when experimental ratio of values for two methods is also the smallest.

It has to be noted that the $^{127}$I method has two main inefficiencies: 1) only ~30% of (n,γ) reactions results in 137.8 keV isomeric state; 2) the delayed time window starting from 900 ns (the limitation coming from the electronics and PMT after pulses [8]) while the isomeric state decay time is 845 ns [12]. Altogether only 1 in 6 neutrons is detected by the method.

**4. Detector sensitivity and MC**

Detector sensitivity (the Li-6 method) of the NaIL detector to thermal neutrons can be established experimentally by comparison of its counting rate with a calibrated He-3 counter. From our measurements of ambient neutrons simultaneously performed with NaIL and CHM-57 [13] detectors in absence of any shields the sensitivity is $83.2\pm0.9^{stat}\pm6.7^{sys}$ counts per sec for neutron flux of 1 n cm$^{-2}$ sec$^{-1}$, where systematic error is due to uncertainty in the CHM-57 sensitivity. These measurements of ambient neutrons of a level of $10^{-3}$ n cm$^{-2}$ sec$^{-1}$ demonstrated the ability of the NaIL detector to measure low intensity neutron fluxes.

Geant4 [14] MC was performed to estimate sensitivity of the NaIL detector to epithermal neutrons. The neutron spectrum for PuBe source was taken from [15]. An additional systematic uncertainty due to difference between PuBe sources can be significant. Obtained space density of the neutron field (n cm$^{-2}$ sec$^{-1}$) is shown in Figure 6. MC prediction is 3.015 $^{127}$I(n,γ)$^{128}$I reactions per sec in the NaIL detector during the PuBe run. This value is in a surprising agreement with experimentally observed 0.56±0.01 Hz after correction of this number for the mentioned above factor of 6 (efficiency described in the end of section 3) and with taking into account unknown uncertainty of the PuBe neutron energy spectrum. The same MC calculations predict that number of inelastic scatterings of fast neutrons on I-127 is 34.6 per sec. This leads to a possibility of fast neutron detection due to excitation of the 57.6 keV state. Due to high counting rate of single events this reaction was not studied experimentally in our work, but has a potential to be used in future low background measurements as was demonstrated by [11].

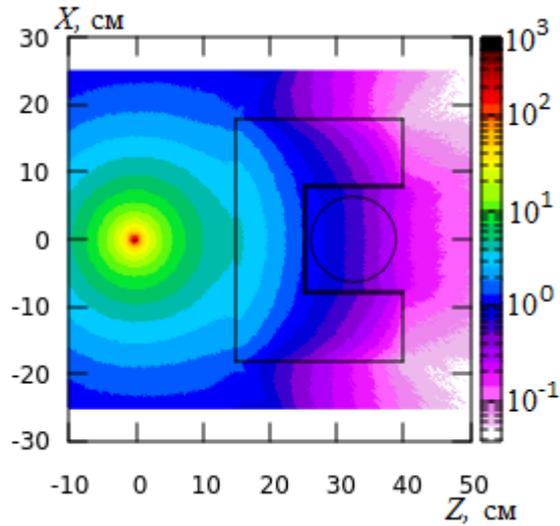

**Figure 6.** Model of neutron field in counts per sec from the PuBe source.

To understand ratio between different branches of neutron's interactions on different depth of the NaIL detector we performed MC of mono energetic parallel neutron beam entering the detector from its top. Several energies in the range from 0.025 eV to 1 MeV were considered. Three of the cases are shown in Figure 7.

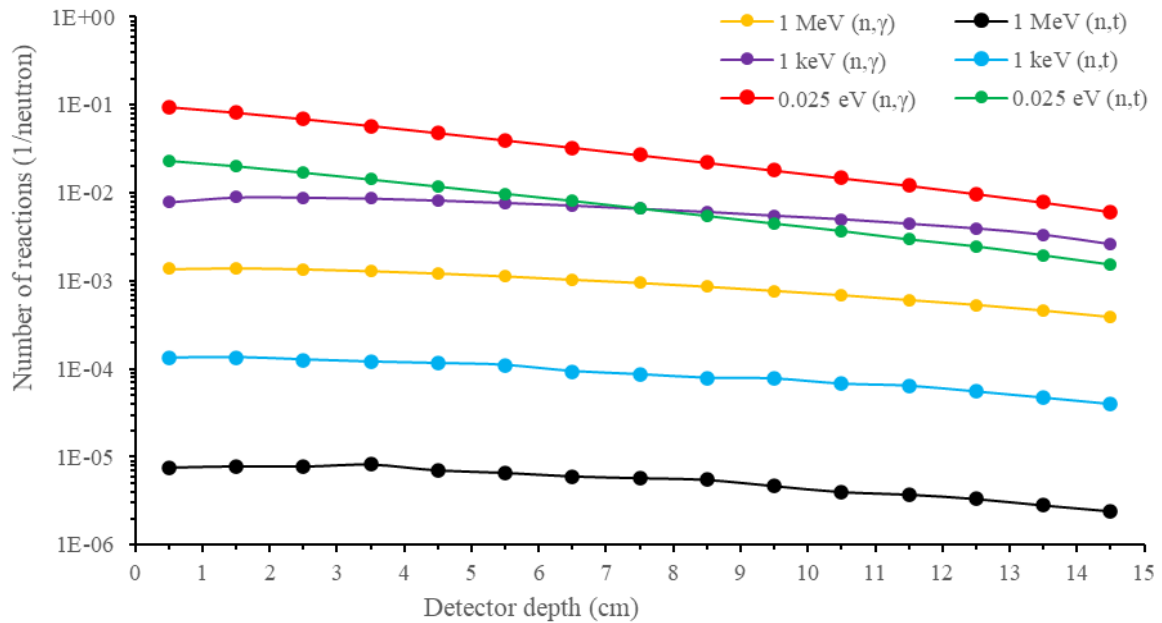

**Figure 7.** Number of interactions per one entering neutron as a function of depth for the NaIL detector with lithium concentration of 1% per mol (Monte Carlo).

Calculations reveal how neutrons of different energies can be detected. According to these calculations, the reaction $^{127}$I(n,γ)$^{128}$I prevails for all neutron energies. Thus, neutrons with different energies will be absorbed mainly by iodine, while $^{6}$Li(n,t)$^{4}$He takes about 20% of all thermal neutrons captured in the detector. It should be emphasized that this fact is crucially important for interpretation of neutron detection with the Li-6 method. Contribution of non-thermal neutrons to $^6$Li(n,t)$^4$He reaction is almost negligible.

The same calculations were performed for a NaIL with 2% per mol concentration of enriched lithium (enrichment by lithium-6 is 95%). Results of these calculations depicted in Figure 8.

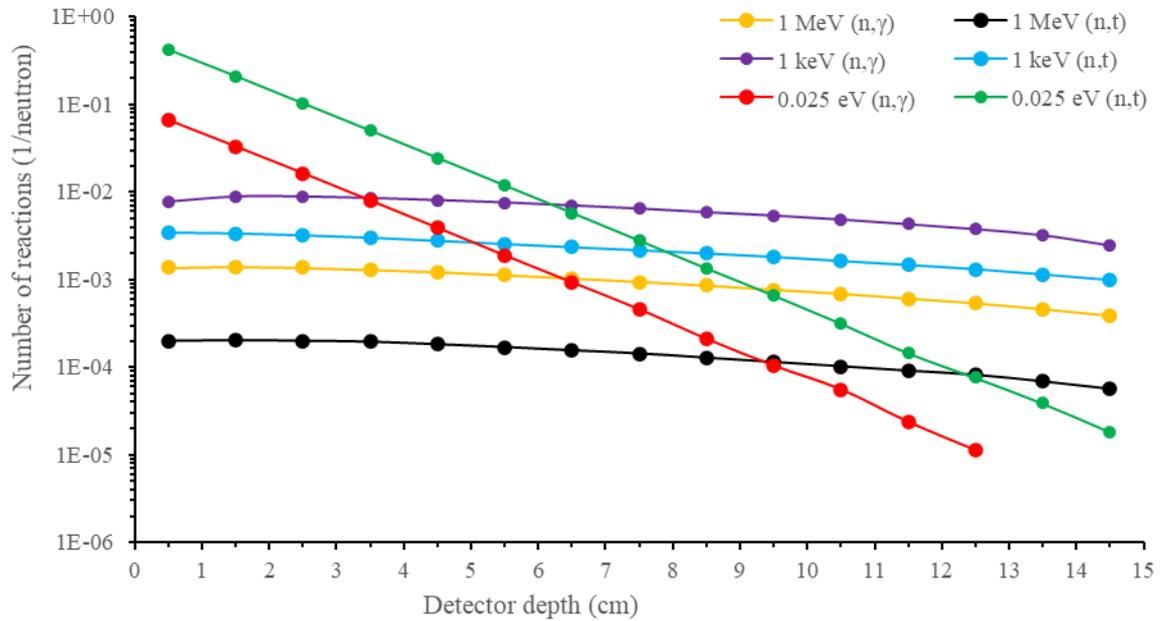

**Figure 8.** Number of interactions per one entering neutron as a function of depth for the NaIL detector with 2% of enriched lithium per mol (Monte Carlo).

With 2% of enriched lithium the situation for neutrons with energies higher than 0.025eV is the same as above. But in the case of thermal neutrons the difference is evident (compare Figures 7 and 8). Now about 85% of all thermal neutrons are absorbed in the detector by $^6$Li(n,t)$^4$He reaction. Further enhancement of lithium-6 concentration in a NaIL detector will naturally increase the number of $^6$Li(n,t)$^4$He reactions and reduce thermal neutron captures by the I-127.

A comparison of the performed MC's with experimental data suggest that the real concentration of natural lithium in the used detector is about 2% per mol instead of 1% specified by the manufacturer as the minimal concentration.

In general, as already noted, the lithium works very efficiently for thermal neutrons near the surface of the detector, thus almost no thermal neutrons can be found in the bulk of the detector where other reactions dominate.

## 5. Conclusion

In this work we present experimental investigation of new type of inorganic scintillating detector containing lithium NaI(Tl+Li). We found that such detector together with its usual use for γ– measurements, can be efficiently applied for detection of neutrons of different energies. Three main reactions for that are $^6$Li(n,t)$^4$He (thermal neutrons), $^{127}$I(n,γ)$^{128}$I (epithermal neutrons), $^{127}$I(n,n')$^{127}$I* (fast neutrons). As different reactions prevail at different depths, several detectors of different sized can be applied for experimental determination of neutron spectrum even in the case when its model is unknown. The studied detector reveals its contamination with

α– emitters. For low background experiments, where neutron measurements are in demand, purified materials should be used.

## Acknowledgments

This work is supported by the Ministry of science and higher education of the Russian Federation under the contract No. 075-15-2020-778 in the framework of the Large scientific projects program within the national project "Science".